\documentclass[conference]{IEEEtran}
\IEEEoverridecommandlockouts
\usepackage{geometry}
\usepackage{graphicx}
\usepackage{subcaption}
\usepackage{booktabs}
\usepackage{cite}
\usepackage{amsmath,amssymb,amsfonts}
\usepackage{algorithmic}
\usepackage{textcomp}
\usepackage{xcolor}
\usepackage{url}
\def\BibTeX{{\rm B\kern-.05em{\sc i\kern-.025em b}\kern-.08em
    T\kern-.1667em\lower.7ex\hbox{E}\kern-.125emX}}

\begin{document}

\title{Performance of Machine Learning Classification in Sonomammogram Images Using BI-RADS\\}

\author{\IEEEauthorblockN{Malitha Gunawardhana}
\IEEEauthorblockA{\textit{Auckland Bioengineering Institute} \\
\textit{University of Auckland}\\
New Zealand \\
}
\and
\IEEEauthorblockN{Norbert Zolek}
\IEEEauthorblockA{\textit{Institute of Fundamental Technological Research} \\
\textit{Polish Academy of Sciences}\\
Poland\\
}
}

\maketitle

\begin{abstract}
This research aims to investigate the classification accuracy of various state-of-the-art image classification models across different categories of breast ultrasound images, as defined by the Breast Imaging Reporting and Data System (BI-RADS). The experiments used 2,945 sonomammogram images for training and 936 images for validation from a cohort of 1,540 patients, with zero patient overlap between the two sets. In order to conduct a thorough analysis, we evaluated six classification architecture families, comprising seven model variants: VGG19 \cite{simonyan2014very}, ResNet50 \cite{he2016deep}, GoogLeNet \cite{szegedy2015going}, ConvNeXt \cite{liu2022convnet}, EfficientNet-V2 \cite{tan2021efficientnetv2}, and two Vision Transformer (ViT) variants \cite{dosovitskiy2020image}. We evaluated the models in three different settings: full fine-tuning, linear evaluation, and training from scratch. Full fine-tuning achieved the strongest overall performance, with ConvNeXt attaining an accuracy of 76.39\% and an F1 score of 67.94\%. These findings demonstrate the potential of transfer learning and contemporary image classification architectures for BI-RADS-based analysis of sonomammograms.
\end{abstract}


\section{Introduction}

Ultrasound imaging is an important tool in breast imaging and supports the detection and diagnostic evaluation of breast abnormalities \cite{jalalian2013computer}, \cite{moon2013computer}. In this work, the term \emph{sonomammogram} refers to a two-dimensional ultrasound image of the breast, and \emph{sonomammography} refers to breast imaging performed using ultrasound. This non-invasive imaging modality employs high-frequency sound waves to create detailed images of breast tissue, providing information that is useful for the characterization of breast masses. The ability of ultrasound to provide real-time imaging also facilitates minimally invasive procedures such as fine-needle aspiration or core-needle biopsy under ultrasound guidance.

Breast cancer remains a critical global health issue, with its incidence and impact spanning across nations and demographics. The World Health Organization (WHO) reported that in the year 2020, approximately 2.3 million women were diagnosed with breast cancer, leading to a devastating toll of 685,000 fatalities worldwide \cite{who2023}. Notably, by the end of the same year, 7.8 million women diagnosed with this malignancy within the preceding five years were still battling the disease. This underscores breast cancer's prominence as the most prevalent form of cancer globally, affecting women across all countries post-puberty, with the risk notably increasing with age. Although predominantly a female disease, it is important to acknowledge that breast cancer does not exclusively affect women; men are also susceptible, albeit at a significantly lower rate. It is well-established that early diagnosis, coupled with consistent and comprehensive treatment, markedly improves patient outcomes and enhances the tolerability of interventions.

The Breast Imaging Reporting and Data System (BI-RADS), developed by the American College of Radiology, provides a standardized framework for categorizing breast imaging findings and guiding subsequent clinical management. The BI-RADS framework uses seven main assessment categories, numbered from 0 to 6. Category 0 indicates an incomplete assessment requiring additional imaging, whereas Category 6 indicates a known biopsy-proven malignancy. Category 4 is further subdivided into 4A, 4B, and 4C, reflecting increasing levels of suspicion for malignancy.

This, in turn, fosters consistency in reporting and recommendations, ensuring a unified approach to patient care. Despite its widespread adoption and the crucial role it plays in the clinical workflow, there exists a conspicuous dearth in the exploration of machine learning-based applications tailored to optimize the BI-RADS assessment process. This research endeavour aims to bridge this gap, offering a comprehensive evaluation of the potential synergies between machine learning methodologies and the BI-RADS system.

To this end, we have undertaken an extensive analysis of various classification networks, spanning VGG19 \cite{simonyan2014very}, ResNet50 \cite{he2016deep}, GoogLeNet \cite{szegedy2015going}, ConvNeXt \cite{liu2022convnet}, and EfficientNet-V2 \cite{tan2021efficientnetv2}. Furthermore, we have explored Vision Transformer (ViT) \cite{dosovitskiy2020image} backbones, aiming to provide a broad comparison of contemporary architectures for BI-RADS-based sonomammogram classification.

This study, therefore, evaluates how contemporary machine learning architectures can support the BI-RADS assessment process. By comparing multiple convolutional and transformer-based models under common training strategies, we aim to improve the precision and reliability of computer-aided analysis of breast ultrasound images.

\begin{figure}[!htp]
    \centering
    \includegraphics[width=1\linewidth]{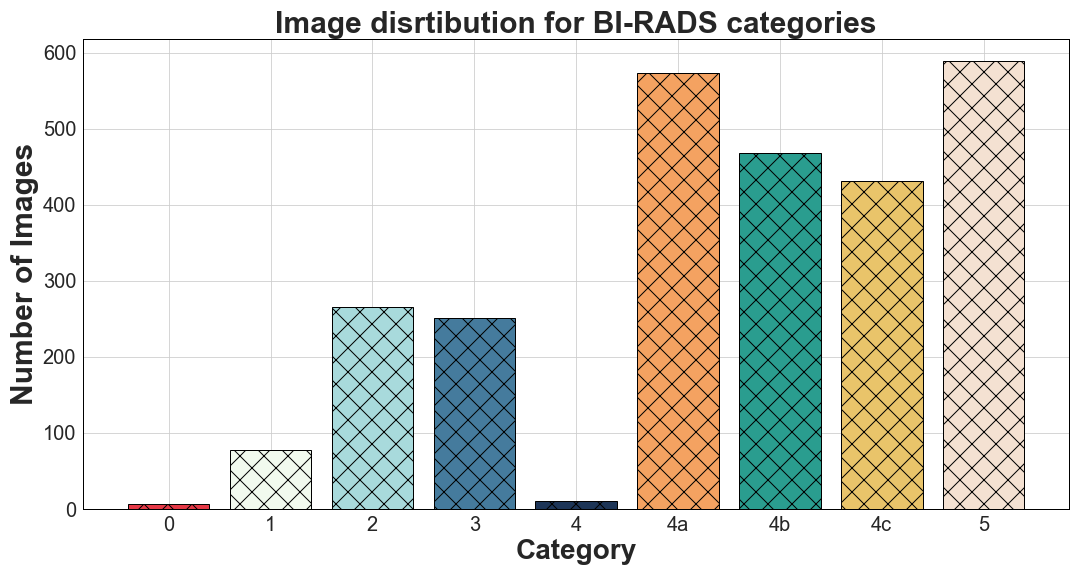}
    \caption{The distribution of images across various BI-RADS categories in the dataset.}
    \label{fig:counts}
\end{figure}

\section{Related Works}

In the domain of Computer-Aided Diagnosis (CAD) systems for breast imaging, numerous efforts have been made to incorporate the influential features from the Breast Imaging Reporting and Data System (BI-RADS). A notable study by Shan et al. (2016) \cite{shan2016computer} delves into this domain, utilizing a dataset comprising 283 images of both benign and malignant lesions. The research focused on harnessing the capabilities of a variety of machine learning algorithms, namely Decision Trees (DT), Artificial Neural Networks (ANN), Random Forest (RF), and Support Vector Machines (SVM), with an aim to enhance the efficiency and accuracy of breast cancer diagnosis. Similarly, Fleury et al. (2019) \cite{fleury2019performance} conducted a comprehensive study employing 206 lesions to meticulously extract radiomic features. Their research extended to the application of an extensive set of five machine learning models: DT, SVM, RF, Multilayer Perceptron (MLP), and Linear Discriminant Analysis (LDA). The goal was to ascertain the most effective algorithm that would result in optimal diagnostic performance. In the study conducted by Chang et al. (2020) \cite{chang2020novel}, the authors employed a tailor-made convolutional neural network (CNN) alongside RF and SVM. These methods were integrated with feature selection techniques to devise a comprehensive multi-class evaluation system for determining the severity of breast tumours, using the BI-RADS as a basis. Interlenghi et al. (2022) \cite{interlenghi2022machine} developed and validated a machine learning model based on radiomics to predict the BI-RADS category of suspicious breast masses detected through ultrasound, with the ultimate goal of decreasing the necessity for biopsies. The model was trained and tested using a dataset comprising 821 ultrasound images representing 834 distinct breast lesions from a cohort of 819 patients, all of whom were initially recommended for ultrasound-guided core needle biopsy due to the suspicious nature of their breast masses.

Prior studies have used conventional machine learning, radiomic features, and custom neural-network approaches for BI-RADS-related breast ultrasound analysis. In contrast, this study focuses on a systematic comparison of established end-to-end image classification architectures under three common training strategies.

\begin{table*}[!htp]
\centering
\caption{BI-RADS classification}
\label{table:birads}
\scalebox{1.2}{
\begin{tabular}{c|p{10cm}}
\toprule
\textbf{Category} & \textbf{Definition} \\ \midrule
0 & Incomplete - Need additional imaging \\
1 & Negative - No masses or suspicious calcifications \\
2 & Benign - No cancer, benign calcification or masses \\
3 & Probably benign - less than 2\% probability of malignancy. Short-interval follow-up is recommended \\
4 & Suspicious abnormality - 2\%-94\% probability of malignancy. \\
4a & Low suspicion for malignancy (2\%-9\%) \\
4b & Moderate suspicion for malignancy (10\%-49\%) \\
4c & High suspicion for malignancy (50\%-94\%) \\
5 & Highly suggestive of malignancy - \textgreater95\% probability of malignancy \\
6 & Known biopsy-proven malignancy \\ \bottomrule
\end{tabular}}
\end{table*}

\section{Methodology}

\subsection{Dataset}

In the course of this research, we analyzed sonomammograms to investigate BI-RADS-derived breast cancer risk categories. The source cohort comprised 1,540 unique patients. The experiments used 2,945 sonomammogram images for training and 936 images for validation. The original images were stored in different formats (JPEG, PNG, and BMP) and at different spatial resolutions. Sonomammograms were acquired using multiple ultrasound systems, including Samsung RS85, Philips, Hitachi ARIETTA 70, Samsung HERA W10, Alpinion XCUBE 90, Voluson E8, and Esaote 6150.

A preliminary examination of the dataset, as illustrated in Figure \ref{fig:counts}, revealed a substantial imbalance across the BI-RADS classes.

\subsection{Data Preprocessing}

\begin{figure*}[!htp]
\centering
\begin{subfigure}{0.49\textwidth}
    \centering
    \includegraphics[width=\linewidth]{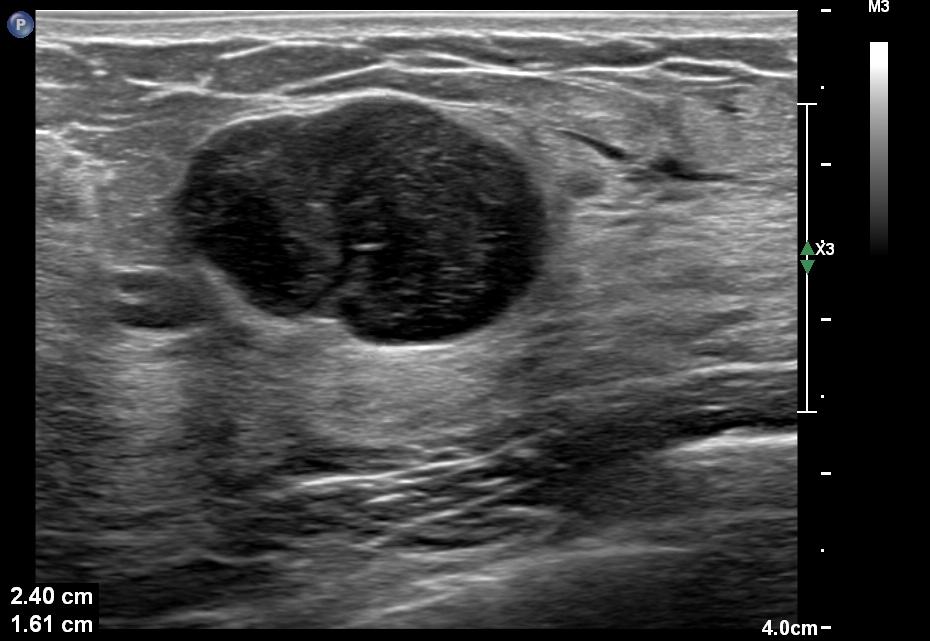}
    \caption{Raw image}
    \label{fig:not_processed}
\end{subfigure}
\begin{subfigure}{0.49\textwidth}
    \centering
    \includegraphics[width=\linewidth]{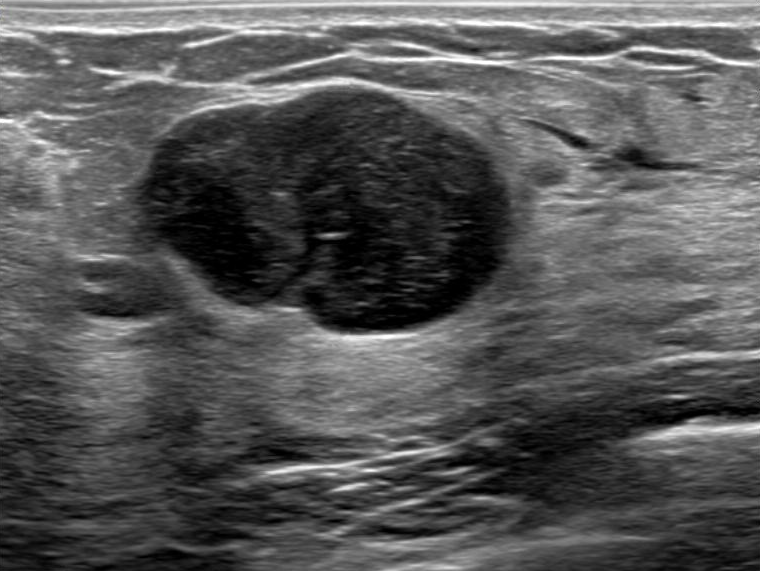}
    \caption{Preprocessed image}
    \label{fig:preprocessed}
\end{subfigure}
\caption{Example of a raw sonomammogram and the corresponding preprocessed image}
\label{fig:processed_images}
\end{figure*}

In the raw images presented in Figure \ref{fig:not_processed}, there exists extraneous data that does not contribute to the primary objective of our analysis. To mitigate this issue and enhance the quality of the dataset, we have initiated a preprocessing procedure. This procedure involves the meticulous cropping of the images to eliminate any superfluous information, ensuring that only the pertinent data is retained. The processed images, exemplified in Figure \ref{fig:preprocessed}, are then subjected to a format conversion, transforming them into the PNG format. This choice of format is intentional, as PNG is renowned for its lossless compression, ensuring that the image quality is preserved. Finally, to maintain uniformity across the dataset and facilitate efficient processing, we resize every image to a resolution of 224$\times$224 pixels. This standardization is crucial as it ensures consistency in the input data for any subsequent analytical or machine-learning models, contributing to the reliability and accuracy of the results obtained.

To mitigate the issues stemming from class imbalance and to facilitate a more efficacious training process, we undertook a systematic reclassification of the dataset. The original categorization comprised nine distinct classes. These were subsequently consolidated into three primary risk categories, ensuring a more balanced representation across the dataset. Specifically, classes 0 through 4 were amalgamated into the 'low-risk' category, classes 4a and 4b were designated as 'moderate-risk', and the remaining classes 4c and 5 were classified as 'high-risk'. This reclassification was introduced to reduce class imbalance and simplify the learning task. These three groups are study-defined labels used for the machine learning experiments and should not be interpreted as replacements for the standard clinical BI-RADS assessment categories.

\subsection{Training and Validation Sets}

Post-reclassification and preprocessing, the data were divided into training and validation sets, with 2,945 images allocated to training and 936 images reserved for validation. A crucial aspect of this partitioning was ensuring zero patient overlap between the two sets. This patient-level separation was used to reduce information leakage and bias arising from patient-specific characteristics.

\subsection{Model Selection and Training}

Our model selection was guided by the objective of assessing and comparing the performance of various established classification architectures. To this end, we selected a range of models, including VGG19 \cite{simonyan2014very}, ResNet50 \cite{he2016deep}, GoogLeNet \cite{szegedy2015going}, ConvNeXt \cite{liu2022convnet}, EfficientNet-V2 \cite{tan2021efficientnetv2}, and ViT \cite{dosovitskiy2020image}. Each of these models represents a unique approach to image classification, providing a comprehensive perspective on their applicability to sonomammogram image analysis. All models were implemented using the PyTorch framework \cite{paszke2019pytorch}.

\subsubsection{Training Approaches}

We employed three distinct training approaches for each of the selected models: fine-tuning using ImageNet pre-trained weights, linear probing with ImageNet weights, and training the models from scratch. Fine-tuning is a more complex and computationally expensive approach which involves training the entire pre-trained model on the new task's labelled data. Linear evaluation is a relatively simple and efficient approach that involves training a new classifier on top of the pre-trained model's feature extractor. The rationale behind using ImageNet weights was to leverage the rich feature representations learned by these models on a diverse set of images, potentially enhancing their performance on our sonomammogram dataset.

\subsubsection{Performance Evaluation}

Model performance was evaluated by comparing the three training approaches using accuracy and F1 score. Full fine-tuning of ImageNet-pretrained models produced the strongest results across all evaluated architectures, followed by linear evaluation and then training from scratch.

\begin{table*}[ht]
\centering
\caption{Comparison of Model Accuracies and F1 Scores across Training Methods}
\label{tab:model_comparison}
\scalebox{1.2}{
\begin{tabular}{l|c|c|c|c|c|c}
\hline
Model & \multicolumn{2}{c|}{FFT} & \multicolumn{2}{c|}{linear} & \multicolumn{2}{c}{Scratch} \\
\cline{2-7}
 & Accuracy & F1 & Accuracy & F1 & Accuracy & F1 \\
\hline
ConvNeXt & \textbf{76.39} & \textbf{67.94} & \textbf{67.40} & \textbf{55.85} & \textbf{61.22} & \textbf{49.55} \\
EfficientNet-V2 & 73.83 & 64.06 & 63.78 & 53.11 & 57.78 & 47.88 \\
GoogLeNet & 72.14 & 62.54 & 62.39 & 52.36 & 54.36 & 44.86 \\
ResNet50 & 69.87 & 60.39 & 60.15 & 49.43 & 51.02 & 44.12 \\
VGG19 & 72.65 & 63.43 & 63.54 & 51.28 & 58.44 & 47.13 \\
ViT(b\_16) & 73.94 & 64.30 & 64.26 & 54.31 & 56.85 & 48.37 \\
ViT(l\_32) & 73.29 & 64.91 & 64.16 & 52.89 & 56.17 & 46.23 \\
\hline
\end{tabular}}
\end{table*}

\section{Results Analysis}

In this section, we present the results of the evaluated methods. Table \ref{tab:model_comparison} provides a comprehensive comparison of the effectiveness of the selected seven model variants, evaluated under three different training paradigms: full fine-tuning (FFT), linear evaluation, and training from scratch. We have meticulously assessed their performance using two crucial metrics: Accuracy and the F1 score, to derive a holistic understanding of their capabilities.

Under full fine-tuning, model accuracies ranged from 69.87\% to 76.39\%, with F1 scores ranging from 60.39\% to 67.94\%. Linear evaluation produced lower accuracies, ranging from 60.15\% to 67.40\%, and F1 scores from 49.43\% to 55.85\%. Training from scratch produced the lowest overall performance, with accuracies from 51.02\% to 61.22\% and F1 scores from 44.12\% to 49.55\%. Thus, the results consistently followed the pattern full fine-tuning $>$ linear evaluation $>$ training from scratch.

ConvNeXt achieved the highest accuracy and F1 score under all three training strategies. Under full fine-tuning, it reached 76.39\% accuracy and a 67.94\% F1 score. For the Vision Transformer variants, ViT(b\_16) achieved slightly higher accuracy than ViT(l\_32) across all three training settings, while their F1 scores were comparable and varied by training strategy.

\section{Conclusion, Constraints, and Directions for Future Research}

This research embarked on a comprehensive examination of various machine learning classification architectures, applying them across three distinct evaluative scenarios: full fine-tuning, linear evaluation and training from scratch. Our findings contribute significantly to the existing body of knowledge, yet there are certain avenues that merit further exploration.

Evaluating additional architectures could provide a broader understanding of performance differences across model families. Furthermore, the current study did not delve into the potential biases that might arise from utilizing different ultrasound devices, a factor that could have a non-negligible impact on the results.

In our methodology, we made a conscious decision to consolidate the BI-RADS categories into three broader classes, a move necessitated by the need to mitigate the challenges posed by data imbalance. However, we acknowledge that this reclassification may have introduced its own set of limitations, potentially masking nuances that are present in the original categorization.

As we look ahead, our aspiration is to amass a larger dataset that retains the original BI-RADS categories. By doing so, we aim to conduct a more granular analysis that preserves the inherent complexity of the data, potentially uncovering insights that were not accessible in the current study. In doing this, we hope to not only validate the findings of this research but also to expand upon them, providing a more comprehensive and nuanced understanding of the machine learning classification methods in the context of ultrasound image analysis.

\end{document}